\documentstyle[12pt,psfig]{article}
\def\appendix{\par
  \setcounter{section}{0}
  \setcounter{subsection}{0}
  \def\@chapapp{\appendixname}
  \def\thesection{\Alph{section}}}
\newcommand{\arctg}{\mathop{\rm arctg}\nolimits}

\begin{document}
\author{Zhmudsky A.A.}
\date{\today}
\title{One class of integrals evaluation in magnet solitons theory}
\maketitle
\begin{abstract}
An analytical-numeric calculation method of extremely  complicated
integrals is presented. These integrals appear often in magnet soliton
theory.

The appropriate analytical continuation and a corresponding integration
contour allow to reduce the calculation of wide class of integrals to a
numeric search of integrand denominator roots (in a complex plane) and a
subsequent residue calculations. The acceleration of series convergence of
residue sum allows to reach the high relative accuracy limited only by
roundoff error in case when $10\div 15$ terms are taken into account.

The circumscribed algorithm is realized in the C program and tested on the
example allowing analytical solution. The program was also used to
calculate some typical integrals that can not be expressed through
elementary functions. In this case the control of calculation accuracy was
made by means of one-dimensional numerical integration procedure.
\end{abstract}
\section{Introduction}
Nonlinear excitations (topological solitons \cite{KIK}) play an important
role in the physics of low-dimensional magnets \cite{BI,IvK}. They
contribute greatly to a heat capacity, susceptibility, scattering
cross-section and other physical characteristics of magnets. In particular,
for two-dimensional (2D) magnets with discrete degeneracy it is important
to take into account the localized stable (with quite-long life time) 2D
solitons \cite{BI,IvK}. According to the experiments \cite{W}, these
solitons determine the relaxation of magnetic disturbance and can produce
peaks in the response functions.

The traditional model describes the magnet state in terms of the unit
magnetization vector $\vec m$, $\vec m^2=1$ with the energy function in the
form
\begin{equation}
W=\int\limits_{}^{}d^2x\left\{A(\nabla\vec m)^2+w_0(\vec m)\right\},
\label{resid:1}\end{equation}
where A is the nonuniform exchange constant and $w_0(\vec m)$ - the
anisotropy energy.

In an anisotropic case the solution is multi-dimensional, that is why the
soliton structure is determined by the system of equations in partial
derivatives. There are no general methods of finding the localized
solutions to such equations and analyzing the stability of the solution.
For this reason a direct variational method is often used. Consequently, a
choice  of a trial function plays a key role in such analysis
\cite{ZyI,IZ,Steph}. In most cases a quite successful choice of trial
function is as follows:
\begin{equation} tg{\theta\over 2}={R\over r}\exp\left(-\frac ra\right)
(1+C_1\cos2\chi),\qquad \varphi=\chi+C_2\sin2\chi+\varphi_0,
\label{resid:2}\end{equation}
where $r$, $\chi$ are the polar coordinates in a magnetic plane, $R$, $a$,
$C_1$, $C_2$ and $\varphi_0$ - variated parameters.

In papers \cite{ZyI,IZ,Steph} the iterative Newton method of solving the
non-linear system of algebraic equations \cite{Press} was used to find the
variated parameters providing an energy minimum. The numerical algorithm
mentioned above results in the necessity of multiple calculation of
two-dimensional integrals (by the angle $\chi$ and radius $r$). Experience
reveals that the calculation process could be essentially accelerated (and
the calculation precision improved) if the analytical-numeric procedure for
the integrals by radius are made beforehand. The typical form of these
integrals is:
\begin{equation}
\int\limits_0^{\infty}{f(r)\exp(mr)dr\over\left[(r\exp(r))^2+a^2\right]^n}
\label{resid:3}\end{equation}
where $n,m$ - integer numbers ($m<2n$), $f(r)$ - polynomial of degree $k$
in variable $r$. It is convenient to denote $g(r)\equiv f(r)\exp(mr)$.
Possible expressions for $g(r)$ are $r\exp(r)$, $(1+r)\exp(r)$,
$r^2\exp(r)$ and so on. Note, that $a$ is essential parameter - no
substitution exists to eliminate it. Below we shall consider the case
$n=1$. From the one hand it is, practically, always possible to reduce the
calculation at $n>1$ to a sum of integrals at $n=1$. From the other hand it
is not difficult to modify evaluation scheme if $n>1$.

Further we show that the indicated type of integrals may be calculated both
analytically and numerically (by the use of joint analytical and numerical
methods). To obtain the desired accuracy one should find integrand residues
(numerically) and build approximate expression (see below).

Main analytical expressions that allow to find integrals like
(\ref{resid:3}) are presented in Analytical Formulae and the C version of
corresponding program is given in Program Realization.

The theory of function of a complex variable gives the powerful method of
definite integrals calculation. In the considered case the algorithm
realization is quite problematical because the expressions for the roots in
a complex plane could not be written analytically. Correspondingly, one can
not write down and analytically summarize expressions for residues.

Present paper pays attention to the fact that numerical methods (roots
search in complex plane and calculation of residue sum) together with
analytical ones allow to use the theory of complex variable with the same
efficiency like in the traditional consideration.

\section{Analytical formulae}
The usual way of integral (\ref{resid:3}) evaluation is to continue
analytically the integrand function in some complex domain $D$. In this
case the evaluating integral will be the part of the integral over the
closed contour $C$ in a complex plane. This contour comprises the real axis
interval and the arc $S$ closing the integration contour. So, the solution
of the problem can be readily obtained if integral value over arc $S$ tends
to the zero. The integral value over contour $C$ can be calculated with a
help of the Residue Theorem. In some cases the evaluating integral will be
the real or imaginary part of a contour integral.

Let's consider the function of a complex variable
\begin{equation} F(z)={g(z)ln(z)\over\ (z\exp(z))^2+a^2},
\label{resid:4}\end{equation}
For this integrand function the conditions of Cauchy's theorem and Jordan's
lemma are fulfilled. So the integral along the arc of infinite radius is
vanishing and $g(z)$ have no branch point. Note that all the examples above
meet these conditions.

\subsection{Integration contour}
\begin{figure}[h]
\begin{center}
\mbox{\psfig{figure=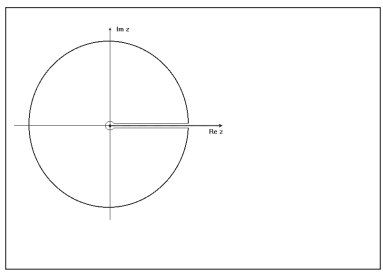,width=50mm}}
\end{center}
\caption{\label{fig1}Integration contour}
\end{figure}

Since $\ln(z)$ is a multiple-valued function, it is necessary to use the
cut plane. Usual method (see, for example \cite{Lavr}) of dealing with
integrals of this type is to use a contour large circle $C_{\infty}$,
center the origin, and radius $R$; but we must cut the plane along the real
axis from $0$ to $\infty$ and also enclose the branch point $z=0$ in a
small circle $c_0$ of radius $r$. The contour is illustrated in
Fig~\ref{fig1}.

Evidently, we may write down the Cauchy's theorem in a form:
\begin{eqnarray}
\int\limits_0^{\infty}{g(x)\ln(x)dx\over (x\exp(x))^2+a^2}+
\int\limits_{C_{\infty}}^{}{g(z)\ln(z)dz\over (z\exp(z))^2+a^2}+
\int\limits^0_{\infty}{g(x)(\ln(x)+2\pi i)dx\over (x\exp(x))^2+a^2}
\nonumber\\
+\int\limits_{c_0}^{}{g(z)\ln(z)dz\over (z\exp(z))^2+a^2}=
2\pi i\sum\limits_{k=0}^{\infty} res F(z_k),
\label{residue:1}\end{eqnarray}
where $C_{\infty}$ is an arc of infinitely great radius, $c_0$ - an arc of
an infinitesimally one. Integrals over these circles are vanishing. Also
taking into account the cancellation of integrals evaluated in opposite
directions we obtain:
\begin{equation}
\int\limits_0^{\infty}{g(x)dx\over (x\exp(x))^2+a^2}=
-\sum\limits_{k=0}^{\infty}res F(z_k).
\label{residue:2}\end{equation}

Thus the evaluating integral is equal to a sum of integrand residues inside
the contour. The direct calculation of the residue sum in integrand poles
leads to extremely slow convergence of the partial sums. Nevertheless, we
shall see below that the convergence series acceleration (Euler's
transformation \cite{Korn}) allows to take at most $10\div 15$ terms into
account.

Following the obvious method one must search the denominator roots and
substitute the evaluated residues in series (\ref{residue:2}).

\subsection{Denominator zeros}
The denominator zeros are roots of two equations:
\begin{equation}
z\exp(z)-ia=0,\quad \mbox{and}\quad  z\exp(z)+ia=0.
\label{resid:5}\end{equation}
Further we will consider only the first equation of (\ref{resid:5}) as the
solution of the second one is complex conjugate to the first one.
Separating real ($x=Re z$) and imaginary ($y=Jm z$) parts of equation
(\ref{resid:5}) leads to a set (system) of two nonlinear equations:
\begin{equation}
x\cos y-y\sin y=0,\qquad y\cos y+x\sin y-a\exp(-x)=0.
\label{resid:7}\end{equation}
Simple algebraic transformations allow to reduce this system to:
\begin{equation}
\sin y={x\exp(x)\over a},\qquad x\cos y-y\sin y=0
\label{resid:8}\end{equation}
As it will be shown below some roots of equation (\ref{resid:5}) have the
positive real part ($x>0$). Nevertheless, in this case the magnitude
$x\exp(x)/a$ does not exceed unit. For the roots with negative real part
($x<0$) at $a>1/e$ always $|x\exp(x)/a|<1$. At $a<1/e$ there exists a
region of negative $x$ values where $x\exp(x)/a<-1$. But it's easy to show
that no roots of (\ref{resid:5}) meet this region.

Let's write down the solution of the first equation (\ref{resid:8}) in a form:
\begin{equation}
y=(-1)^N \arcsin{x\exp(x)\over a}+N\pi,\qquad N=0,\pm 1,\pm 2,\pm 3,\ldots.
\label{resid:9}\end{equation}
Integer number $N$ separate different branches of function $\arctg$. The $y$
value for these branches lie in the limits:
\begin{equation} (2N-1)\frac {\pi}2\le y\le(2N+1)\frac {\pi}2,
\quad N=0,\pm 1,\pm 2,\pm 3,\ldots.
\label{resid:6}\end{equation}

After substitution expression (\ref{resid:9}) into the second equation of
the system (\ref{resid:7}), we obtain:
\begin{equation}   (-1)^N\arcsin{x\exp(x)\over a}+N\pi
=\pm a\exp(-x)\sqrt{1-\left({x\exp(x)\over a}\right)^2}.
\label{resid:10}\end{equation}

If $N=\pm 2m$ ($m=0,1,2,\ldots$) $\cos y>0$ and we chose the upper sign
(plus) in the right-hand side of equation (\ref{resid:10}). Contrary to
this, $\cos y<0$ if $N=\pm(2m+1)$ and we must chose the lower sign (minus)
in the right-hand side of equation (\ref{resid:10}). Thus at even $N$
($N=\pm 2m$) we obtain:
\begin{equation}
\arcsin{x\exp(x)\over a}+N\pi=a\exp(-x)\sqrt{1-\left({x\exp(x)\over
a}\right)^2},        \label{re:6}\end{equation}
and at odd $N$ ($N=\pm (2m+1)$) respectively:
\begin{equation}
\arcsin{x\exp(x)\over a}-N\pi=a\exp(-x)\sqrt{1-\left({x\exp(x)\over
a}\right)^2}.        \label{re:7}\end{equation}

It is easy to see, that equation (\ref{re:6}) have solutions only at $N\ge
0$, while (\ref{re:7}) one only at $N<0$. Therefore, it is convenient to
unite (\ref{re:6}) and (\ref{re:7}) and write down:
\begin{equation}
\arcsin{x\exp(x)\over a}+k\pi-a\exp(-x)\sqrt{1-\left({x\exp(x)
\over a}\right)^2}=0, \quad k=0,1,2,\ldots
\label{re:8}\end{equation}

The point to keep in mind, that $\cos y>0$ at $k=2m$ and $\cos y<0$ at
$k=2m+1$. At even $k=2m$ and $x<0$ it is not difficult to build an
approximation for the real and imaginary parts of the root $z_k$:
\begin{equation} x_k\approx-\left[1+{\ln{k\pi/ a}\over 2(k\pi)^2}\right]
\ln{k\pi\over a},    \qquad y_k\approx k\pi.
\label{residu:10}\end{equation}
As $x<0$ and $y>0$ the argument of the complex number is placed in the
second quarter (less than $\pi$, but greater than $\pi/2$) and is equal to:
\begin{equation}
\arg z_k\approx{\pi\over 2}+\arctg {k\pi\over \ln(k\pi/a)}.
\label{residu:11}\end{equation}
Just note that complex conjugate number argument is placed in the third
quadrant ($x<0$ and $y<0$) and:
\begin{equation}
\arg\overline z_k\approx{3\pi\over 2}-\arctg {k\pi\over \ln(k\pi/a)}.
\label{residu:12}\end{equation}
The comparison of these solutions with exact numerical one will be given
below in Table~\ref{tab:number1}.

\begin{figure}[h]
\begin{center}
\mbox{\psfig{figure=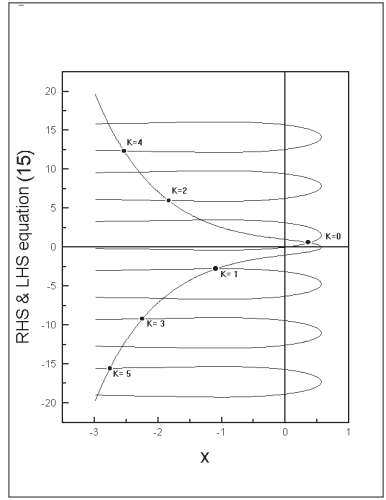,width=50mm}}
\end{center}
\caption{\label{fig2}Denominator zeroes in complex plane}
\end{figure}

At odd $k=2m+1>0$ the argument of the root and complex conjugate value are
equal respectively:
\begin{eqnarray}
\arg z_k\approx{3\pi\over 2}-\arctg {k\pi\over \ln(k\pi/a)}\nonumber\\
\arg \overline z_k\approx{\pi\over 2}+\arctg {k\pi\over \ln(k\pi/a)}.
\label{residu:14}\end{eqnarray}
The comparison of these solutions with exact numerical one will also be
given in Table~\ref{tab:number1}.

Hence, evaluation of the complex roots of equation (\ref{resid:5}) is
reduced to a real part $z_k$ search (transcendental equation (\ref{re:8})
solution) and further calculation of imaginary part $z_k$ by
(\ref{resid:9}). The graphs of the sum of the first two terms of
(\ref{re:8}) and the third term of the equation (\ref{re:8}) are presented
on Fig~\ref{fig2}.

Intersection points of these curves marked by circles correspond to the
roots of equation (\ref{re:8}). Any of the roots marked on Fig~\ref{fig2}.
can be evaluated numerically with the help of Newton-Rafson method (e.g.
\cite{McCraken}). The approximations (\ref{residu:10}) are used as an
initial guess. Necessary explanation will be made in the Program
Realization. It is convenient to number these roots by corresponding values
of $k$ as it is shown on the figure (the numeration is used in the
program).

\begin{table}[th]
\centering
\label{tab:number1}\caption{Relative error between exact value and
approximation \newline
\centerline{for different $a$}}\vspace{2mm}
\begin{tabular}{|r|c|c|c|}
\hline {k}&{a=1/e}&{a=1.0}&{a=10}\\ \hline
$   1  $ & $ 9.21835\cdot 10^{-2} $ &  $ 1.11473\cdot 10^{-1}   $ & $
                                                          $ \\  \hline
$   2  $ & $ 2.40163\cdot 10^{-2} $ &  $ 2.52883\cdot 10^{-2}   $ & $
                                                          $ \\  \hline
$   3  $ & $ 1.09204\cdot 10^{-2} $ &  $ 1.11533\cdot 10^{-2}   $ & $
                                                          $ \\  \hline
$   4  $ & $ 6.21187\cdot 10^{-3} $ &  $ 6.27698\cdot 10^{-3}   $ & $
                                     6.35174\cdot 10^{-3} $ \\  \hline
$   5  $ & $ 4.00030\cdot 10^{-3} $ &  $ 4.02305\cdot 10^{-3}   $ & $
                                     4.06565\cdot 10^{-3} $ \\  \hline
$   6  $ & $ 2.78844\cdot 10^{-3} $ &  $ 2.79747\cdot 10^{-3}  $ & $
                                     2.82172\cdot 10^{-3} $ \\  \hline
$   7  $ & $ 2.05364\cdot 10^{-3} $ &  $ 2.05747\cdot 10^{-3}  $ & $
                                     2.07187\cdot 10^{-3} $ \\  \hline
$   8  $ & $ 1.57493\cdot 10^{-3} $ &  $ 1.57657\cdot 10^{-3}  $ & $
                                     1.58551\cdot 10^{-3} $ \\  \hline
$   9  $ & $ 1.24585\cdot 10^{-3} $ &  $ 1.24651\cdot 10^{-3}  $ & $
                                     1.25229\cdot 10^{-3} $ \\  \hline
$  10  $ & $ 1.01001\cdot 10^{-3} $ &  $ 1.01021\cdot 10^{-3}  $ & $
                                     1.01407\cdot 10^{-3} $ \\  \hline
\end{tabular}
\end{table}

In case $Re z>0$ (it will be if $a>|k\pi|$) it is more convenient to find
roots by bisection because it is difficult to get a necessary initial
guess. At the same time it is obvious from Fig~\ref{fig2} that at $Re z>0$
each subsequent root lies between zero and a root found before.

\subsection{Residue Calculation}
\subsubsection{The test integral}
Let's consider the test example of an integral (\ref{resid:3}) when $m=1$
and $g(r)=(1+r)\exp(r)$. Elementary analytic calculation yields $\frac
12\pi/a$. On the other hand function $F(z)$ residue in the pole of first
order (at the same $g(x)$) is equal:
\begin{equation}
res F(z_k)=\left.{(1+z)\exp(z)\ln(z)\over {\displaystyle{d\over dz}}
[(z\exp(z))^2+a^2]}\right|_{z_k}
=\left.{\ln(z)\over 2z\exp(z)}\right|_{z_k}
=\left.\mp{i\ln(z)\over 2a}\right|_{z_k},
\label{residue:11}\end{equation}
where upper sign (minus) corresponds to the first and lower (plus)  to the
second of the equations (\ref{resid:5}).

Let's take the equation (\ref{residue:2}) in a form:
\begin{eqnarray}
\int\limits_0^{\infty}{(1+x)\exp(x)dx\over (x\exp(x))^2+a^2}=
-\sum\limits_{k=0}^{\infty} res F(z_k)
={i\over 2a}\sum\limits_{k=0}[\ln(z_k)-\ln(\overline z_k)]\nonumber\\
={i\over 2a}\sum\limits_{k=0}[i\arg(z_k)-i\arg(\overline z_k)]
={1\over 2a}\sum\limits_{k=0}[\arg(\overline z_k)-\arg(z_k)].
\label{resid:14}\end{eqnarray}
For the following evaluation it is important to determine root arguments
correctly. Here and below in this article the complex value argument will
be determined by principal value of $\arctg$ in order to use the library
function $\arctg$ within the limits of first quadrant. The point to keep in
mind is that the quadrant of the complex value one have to define "by
hand". It allows to avoid mistakes connected with possible definitions of
library functions of negative argument, etc.

Let's consider first three terms of the series summarized:
\begin{enumerate}
\item If $k=0$ the root $z_0$ is placed in the first quadrant and its'
argument is equal to $\varphi_0=\arctg|y_0/x_0|$. Evidently, that the
complex conjugate value has the argument $2\pi-\varphi_0$. Contribution of
these two poles in the right-hand side of equation (\ref{resid:14}) is:
$$\arg(\overline z_0)-\arg(z_0)=2\pi-2\varphi_0$$
\item The next root $z_1$ (at $k=1$) is placed in the third quadrant and
its' argument is equal to $\pi+\varphi_1=\pi+\arctg|y_1/x_1|$. Complex
conjugate value argument is $\pi-\varphi_1=\pi-\arctg|y_1/x_1|$.
Corresponding contribution of these roots is:
$$\arg(\overline z_1)-\arg(z_1)=-2\varphi_1$$
\item Consider one root more at $K=2$. Its' argument is equal to
$\pi-\varphi_2=\pi- \arctg |y_2/x_2|$. For the complex conjugate value we
obtain $\pi+\varphi_2=\pi+\arctg |y_2/x_2|$. Thus, we find:
$$\arg(\overline z_2)-\arg(z_2)=2\varphi_2$$
\end{enumerate}
Structure of each term is clear that is why we simply quote the result:
\begin{eqnarray}
& &\int\limits_0^{\infty}{(1+x)\exp(x)dx\over (x\exp(x))^2+a^2}
={1\over 2a}\sum\limits_{k=0}[\arg(\overline z_k)-\arg(z_k)]=\nonumber\\
&=&{1\over 2a}\left[2\pi-2\varphi_0-2\varphi_1+2\varphi_2-2\varphi_3+
2\varphi_4-\ldots\right]\nonumber\\
&=&{1\over a}\left[\pi-\left(\varphi_0+\varphi_1-\varphi_2+\varphi_3-
\varphi_4+\ldots\right)\right]
\label{resid:21}\end{eqnarray}
The sum $\varphi_k$ is found numerically. This series is conventional
convergent. Partial sums consecutively equal to $\simeq 0.81$ and $\simeq
2.3$ and converge very slowly. Nevertheless, the program of convergence
acceleration \cite{Press} found the result $\pi/2$ with the relative error
that does not exceed $10^{-8}$ ($10\div 15$ terms of the series are taken
into account).

\subsubsection{First order pole}
In this section we will consider integral (\ref{resid:3}) at $m=1$ and
$g(r)=\exp(r)$. The residue of integrand function $F(z)$ in the first order
pole is evidently equal to:
\begin{equation}
res F(z_k)=\left.{\exp(z)\ln(z)\over {\displaystyle{d\over dz}}
[(z\exp(z))^2+a^2]}\right|_{z_k}
=\left.{\ln(z)\over 2z\exp(z)(1+z)}\right|_{z_k}
=\left.\pm{i\ln(z)\over 2a(1+z)}\right|_{z_k}.
\label{residue:24}\end{equation}
As in the previous case the upper sign (minus) corresponds to the first and
lower (plus) to the second of the equations (\ref{resid:5}).

It is convenient to write down residue expression in a form:
\begin{eqnarray}
\int\limits_0^{\infty}{\exp(x)dx\over (x\exp(x))^2+a^2}=
-\sum\limits_{k=0}^{\infty} res F(z_k)
={i\over 2a}\sum\limits_{k=0}\left[{\ln(z_k)\over 1+z_k}-
{\ln(\overline z_k)\over 1+\overline z_k}\right]\nonumber\\
={i\over 2a}\sum\limits_{k=0}\left[{\ln|z_k|+i\arg(z_k)\over 1+z_k}-
{\ln|\overline z_k|+i\arg(\overline z_k)\over 1+\overline z_k}\right].
\label{resid:25}\end{eqnarray}

The argument value depends on denominator zeros placement. Obviously, there
are four possibilities.

For the complex roots lying in the first quadrant (it will be if
$k=0,2,4\ldots$, at $a>k\pi$) the first term in (\ref{resid:25}) is equal
to:
$${\ln|z_0|+i\varphi_0\over 1+z_0}$$
Second term (for the complex conjugate value) yields:
$${\ln|\overline z_0|+i(2\pi-\varphi_0)\over 1+\overline z_0}$$
Subtracting the terms and separating real and imaginary parts (factor
$i/2a$ is taken into account) we obtain:
\begin{equation}
{y_0\ln|z_0|+(1+x_0)(\pi-\varphi_0)\over a|1+z_0|^2}
+i{\pi y_0\over a|1+z_0|^2}\label{resid:26}\end{equation}
Pay attention that the angle $\varphi_0$ is determined as the principal
value of $\arctg$ in the limits $[0,\pi/2]$.

Similary calculations for the second, third and fourth quadrants give
respectively:
\begin{eqnarray}
{y_2\ln|z_2|+(1+x_2)\varphi_2\over a|1+z_2|^2}+i{\pi y_2\over a|1+z_2|^2}, \\
{y_3\ln|z_3|-(1+x_3)\varphi_3\over a|1+z_3|^2}+i{\pi y_3\over a|1+z_3|^2}, \\
{y_1\ln|z_1|-(1+x_1)(\pi-\varphi_1)\over a|1+z_1|^2}
+i{\pi y_1\over a|1+z_1|^2},
\label{resid:28}\end{eqnarray}

As if structure of each term is already clear then it is not hard to write
the sum of real residue parts (\ref{resid:26}-\ref{resid:28}) and all the
subsequent ones.
\begin{eqnarray}
& &\int\limits_0^{\infty}{\exp(x)dx\over (x\exp(x))^2+a^2}=
-\sum\limits_{k=0}^{\infty}res(F(z_k))\nonumber\\
&=&\frac 1a\left[{y_0\ln|z_0|+(1+x_0)(\pi-\varphi_0)\over |1+z_0|^2}
+{y_1\ln|z_1|-(1+x_1)\varphi_1\over |1+z_1|^2}\right.\nonumber\\
&+&{y_2\ln|z_2|+(1+x_2)\varphi_2\over |1+z_2|^2}
+{y_3\ln|z_3|-(1+x_3)\varphi_3\over |1+z_3|^2}\nonumber\\
&+&\left.{y_4\ln|z_4|+(1+x_4)\varphi_4\over |1+z_4|^2}+\ldots\right]
\label{resid:31}\end{eqnarray}
Sum of not more than 15 terms of this series give right result for any
$a>0.0006$ if acceleration of series convergence is used.  The partial sums
of the imaginary parts also rapidly ($\sim 10\div 15$ terms) tends to the
zero ($\sim 10^{-12}$).

\subsection{Approximation at $a$ vanishing}
At $a$ vanishing it is convenient to write down the approximate expression
for integral (\ref{resid:25}). Let's set $m=1$ in (\ref{resid:3}) and make
substitution:
\begin{equation} x\exp(x)=au, \label{approx:1}\end{equation}
where $u$ is the new independent variable. Left-hand side of equation
(\ref{approx:1}) may be expanded in powers of the x:
\begin{equation} au\approx x+x^2+\frac 1{2!}x^3+\ldots
\label{approx:2}\end{equation}
Inversion \cite{Korn} of this series gives:
\begin{equation} x\approx au-(au)^2+\frac 32(au)^3-\ldots
\label{approx:3}\end{equation}
For the present purposes, it is sufficient to retain only one term in
right-hand side of equation (\ref{approx:3}). We will focus on the case
discussed in the previous section, namely, $f(x)=\exp(x)$. Simple
integration leads to the following result:
\begin{equation} \int\limits_0^{\infty}{\exp(x)dx\over (x\exp(x))^2+a^2}
\approx \frac 1{a(1+a^2)}\left[\frac {\pi}2+a\ln(a)\right]
\label{approx:4}\end{equation}
One more term taking into account yields:
\begin{eqnarray} & &\int\limits_0^{\infty}{\exp(x)dx\over (x\exp(x))^2+a^2}
\nonumber\\ &\approx& \frac 1{(1+a^2)^2+a^2}\left\{(1+a^2)\frac {\pi}{2a}+
\ln(a)+{3+2a^2\over2\sqrt{5}}\ln\left|{1+\sqrt{5}\over \sqrt{5}-1}
\right|\right\}         \label{approx:5}\end{eqnarray}
Quality of these approximations is presented in Table~\ref{tab:number2}.
\begin{table}[htb]
\flushleft
\centering
\caption{Comparison of approximation and exact value of integral}
\label{tab:number2}
\begin{tabular}{|r|c|c|c|}
\hline
$  a  $ & Integral value & relative error for & relative error for \\
        & (exact)  & approximation (\ref{approx:4}) &
                                 approximation (\ref{approx:5}) \\  \hline
$10^{-07}$ & $1.57079\cdot 10^7 $       & $ 3.67\cdot 10^{-8} $ &
                                          $ 4.35\cdot 10^{-9} $ \\  \hline
$10^{-06}$ & $1.57078\cdot 10^6 $       & $ 3.68\cdot 10^{-7} $ &
                                          $ 4.35\cdot 10^{-8} $ \\  \hline
$10^{-05}$ & $1.57069\cdot 10^5 $       & $ 3.67\cdot 10^{-6} $ &
                                          $ 4.35\cdot 10^{-7} $ \\  \hline
$10^{-04}$ & $1.56993\cdot 10^4 $       & $ 3.68\cdot 10^{-5} $ &
                                          $ 4.36\cdot 10^{-6} $ \\  \hline
$6\cdot10^{-04}$ & $2.61115\cdot 10^3 $ & $ 2.21\cdot 10^{-4} $ &
                                          $2.62\cdot 10^{-5}  $ \\  \hline
\end{tabular}
\end{table}
\\[2mm]
As one can see the relative error given by approximation (\ref{approx:5})
is not worth than $2.62\cdot 10^{-5}$ at $a=6\cdot10^{-4}$ and smaller for
other values. If it is necessary, the approximation (\ref{approx:5}) can be
improved.

\section{Program Realization}
Program text consists of the principal routine (main) and eight procedures:
\begin{itemize}
\item Root - subroutine for searching real part of denominator zeros.
Newton-Rafson iteration method used for this purpose.
\item Fun - subroutine for left-hand side equation (\ref{re:8}) evaluation.
\item Der - subroutine for left-hand side of equation (\ref{re:8}) derivative
evaluating.
\item Approximation - evaluate initial guess to the root according to the
(\ref{residu:10}).
\item Eulsum - program of acceleration of series (\ref{resid:31})
convergence. The C version of corresponding FORTRAN program \cite{Press} is
used.
\item Bisection - recursive version of bisection root finding.
\item Right\_bound - auxiliary program. Determine the right bound of root
region.
\item Sign - auxiliary program for Bisection. Determine sign of variable $x$.
\end{itemize}
The demonstration version of the program is presented below. The first four
statements declare the necessary functions that are used in the program.
\vspace{-3mm}
\begin{verbatim}
#include <dos.h>
#include <stdio.h>
#include <math.h>
#include <alloc.h>
\end{verbatim}
\vspace{-3mm}
// Declaration of functions used below.
\vspace{-3mm}
\begin{verbatim}
double Fun(double);
double Der(double);
double Bisection(double& ,double& ,double& ,double&);
double Root(double);
double Approximation_x(void);
double Right_bound(void);
void Eulsum(double&, double, int);
int Sign(double);
\end{verbatim}
\vspace{-3mm}
\begin{tabular}{p{4mm}p{120mm}}
// // // & Defining of the global constants. Parameter $A\equiv a$ can be
defined in any other way. EPS - relative accuracy for root search; K - root
number; N - number of roots taking into account. \\
\end{tabular}
\vspace{-3mm}
\begin{verbatim}
double  A = 0.2, EPS = 1.0e-14;
long unsigned irec = 0, irecmax = 100;
int K, N = 20;
\end{verbatim}
\vspace{-3mm}
\begin{tabular}{p{4mm}p{12cm}}
// // // & Definition of parameters used by procedures Root (x, y, yx,
phi\_k, den, argument) and Bisection (a, c, funa, func, xold, xk). See
below. \\
\end{tabular}
\vspace{-3mm}
\begin{verbatim}
void main(void)
{ double x, y, yx, phi_k, den, argument;
  double real_part = 0, imag_part = 0;
  double a, c, funa, func, xold, xk;
  double *real_term = (double*)calloc(N,sizeof(double));
  double *imag_term = (double*)calloc(N,sizeof(double));
\end{verbatim}
\vspace{-3mm}
\begin{tabular}{p{4mm}p{12cm}}
// //& Two previous statements reserved memory for storaging denominator
roots. Cycle on root numbers begins below.\\
\end{tabular}
\vspace{-3mm}
\begin{verbatim}
  xold = 0.99999*Right_bound();
  for(int k = 0; k < N; k++)
  { K = (k%2) ? -k : k;
\end{verbatim}
\vspace{-2mm}
\begin{tabular}{p{4mm}p{12cm}}
// // // & Variable $K$ (upper case) takes values $0,-1,2-3,\ldots$ while
$k=0,1,2,3,\ldots$. Statements IF ELSE differs three different cases:
$Real(z_k)=0$, $Real(z_k)>0$ and $Real(z_k)<0$.\\
\end{tabular}
\vspace{-3mm}
\begin{verbatim}
    if(A == fabs(K*M_PI)) x = 0.0;
    else { if(A > fabs(K*M_PI))
              { a = 0.0; c = xold;
                funa = Fun(a);   func = Fun(c);
                x = Bisection(a,c,funa,func);
                xold = 0.99999999 * x;  irec = 0;}
            else
              { xk = Approximation_x();
                x = Root(xk);                 }
          }
\end{verbatim}
\vspace{-2mm}
\begin{tabular}{p{4mm}p{12cm}}
// // & The following statements calculate argument of $z_k$ and store real
and imaginary parts of the term to be summarized.  \\
\end{tabular}
\vspace{-3mm}
\begin{tabbing}
\=\hspace{8mm}        \=\hspace{6cm} \\
\> \>{\small\tt       y = pow(-1,k)*asin(x*exp(x)/A) + K*M\_PI; } \\
\> \>{\small\tt       if(x == 0) argument = M\_PI\_2;            } \\
\> \>{\small\tt       else \{  yx = fabs(y/x);                  } \\
\> \>{\small\tt\hspace{12mm}   phi\_k = atan(yx);                } \\
\> \>{\small\tt\hspace{12mm}   argument = (x > 0) ? M\_PI - phi\_k :
                                                               phi\_k;\} } \\
\> \>{\small\tt       den = (1+x)*(1+x)+y*y;                             } \\
\> \>{\small\tt       real\_term[k]=(0.5*y*log(x*x+y*y)+(1+x)*
                                                   Sign(y)*argument)/den;} \\
\> \>{\small\tt       imag\_term[k]=y/den; } \\
\end{tabbing}
\vspace{-5mm}
\verb|  }|

\hspace{-10mm}
\begin{tabular}{p{4mm}p{120mm}}
// // // // & End cycle on root numbers. Accelerating the rate of a
sequence of partial sums performed by the procedure Eulsum. This is the $C$
version of Van Wijngaarden's algorithm (see \cite{Press}). Then the result
is printed. The last two statements free the allocated memory. \\
\end{tabular}
\vspace{-3mm}
\begin{verbatim}
  for(int j = 0; j < N; j++) Eulsum(real_part,real_term[j],j);
  for(    j = 0; j < N; j++) Eulsum(imag_part,imag_term[j],j);
  printf("Real part =%12.5lf ",real_part/A);
  printf("Imaginary part =%12.5le \n",imag_part);
  free(real_term);
  free(imag_term);
}
\end{verbatim}
\vspace{-1mm}
\begin{tabular}{p{4mm}p{120mm}}
// // & Initial guess for the Newton-Rafson iteration is chosen with
respect to parameter $a$ value. The appointments of other statements are
evident.    \\
\end{tabular}
\vspace{-3mm}
\begin{verbatim}
double Root(double xk)
{ double xk1;
  for(int it = 0; it < 30; it++)
   { xk1 = xk - Fun(xk)/Der(xk);
     if(fabs(xk-xk1) <= fabs(xk1)*EPS) break;
     xk = xk1;        }
  return xk1;  }
\end{verbatim}
\vspace{-3mm}
// No comments.
\vspace{-3mm}
\begin{verbatim}
double Fun(double x)
{ double px = x*exp(x)/A, arcsin = asin(px);
  return arcsin + abs(K)*M_PI - A*sqrt(1-px*px)/exp(x);   }
\end{verbatim}
\vspace{-3mm}
// No comments.
\vspace{-3mm}
\begin{verbatim}
double Der(double x)
{ double expa=exp(x), px = A/expa; //px = x*expa/A;
    return (1 + x + px*px)/sqrt(px*px-x*x);   }
\end{verbatim}
\vspace{-3mm}
// Calculation of initial guess according to the expressions
(\ref{residu:10})
\vspace{-3mm}
\begin{verbatim}
double Approximation_x(void)
{ double mu = abs(K)*M_PI, den = mu*mu, lnKPi = log(mu/A);
  if(!K) return 0;
  else return -lnKPi*(1+0.5*lnKPi/den);  }
\end{verbatim}
\vspace{-3mm}
// For details see \cite{Press}.
\vspace{-3mm}
\begin{verbatim}
void Eulsum(double& sum, double term, int jterm)
{ double static wksp[28], dum, tmp;
  int static nterm;
  if(jterm == 0) { nterm=0; wksp[0]=term; sum=0.5*term; }
  else { tmp = wksp[0];  wksp[0] = term;
         for(int j = 0; j < nterm; j++)
           { dum = wksp[j+1];
             wksp[j+1] = 0.5*(wksp[j]+tmp);
             tmp = dum;
              }
         wksp[nterm+1] = 0.5*(wksp[nterm]+tmp);
         if(fabs(wksp[nterm+1]) <= fabs(wksp[nterm]))
           { sum += 0.5*wksp[nterm+1];  nterm += 1; }
        else sum += wksp[nterm+1];
       }
}
\end{verbatim}
\vspace{-3mm}
\noindent
// Usual bisection algorithm realized by means of the recursion.
\vspace{-3mm}
\begin{verbatim}
double Bisection(double& left, double& right,double& fun_left,
                                             double& fun_right)
{ double center = 0.5*(left + right);
  if(++irec < irecmax)
  {if(fabs(left - right) < EPS*fabs(center)) return center;
   double fun_c = Fun(center);
   if(fabs(fun_left-fun_right)<EPS*fabs(fun_c)) return center;
   if(Sign(fun_left) == Sign(fun_c))
                     { left = center ; fun_left = fun_c;}
   else { right = center;  fun_right = fun_c;  }
   center = Bisection(left,right,fun_left,fun_right);
  }
 irec--;
 return center;
}
\end{verbatim}
\noindent
\begin{tabular}{p{4mm}p{120mm}}
// // & Auxiliary program. Determine the right bound of the root region
(see Fig.~\ref{fig2}).
\end{tabular}
\vspace{-3mm}
\begin{verbatim}
double Right_bound(void)
{ double xn, xn1;
  xn = (A > 1) ? log(A) : A;
  for(int k = 0; k < 20; k++)
  { xn1 = xn + (A*exp(-xn) - xn)/(1.0 + xn);
    if(fabs(xn1-xn) <= 1.0e-14*fabs(xn1)) break;
    xn = xn1; }
  return xn1;
}
\end{verbatim}
// Auxiliary program for Bisection. Determine sign of variable $x$.
\begin{verbatim}
int Sign(double x)
{ if(!x) return 0;
  return (x>0) ? 1 : -1;  }
\end{verbatim}

The test result of this programm is presented in Table~\ref{tab:number3}.
\vspace{-3mm}
\begin{table}[h]
\flushleft
\centering
\caption{Exact value of the integral and relative error}\label{tab:number3}
\begin{tabular}{|l|l|c|}
\hline
$\quad  a  $ & \quad \quad Exact & relative error   \\
        & \quad integral value  & for approximation  \\  \hline
$10^{-06}$ & $1.57078308845\cdot 10^7 $    & $ 4.35\cdot 10^{-8} $
                                                              \\  \hline
$10^{-05}$ & $1.57068696938\cdot 10^6 $    & $ 4.35\cdot 10^{-7} $
                                                              \\  \hline
$10^{-04}$ & $1.56993298295\cdot 10^5 $    & $ 4.36\cdot 10^{-6} $
                                                              \\  \hline
$10^{-03}$ & $1.56446267294\cdot 10^7 $    & $ -2.45\cdot 10^{-11} $
                                                              \\  \hline
$10^{-02}$ & $1.53021971263\cdot 10^6 $    & $ \quad 1.51\cdot 10^{-10} $
                                                              \\  \hline
$10^{-01}$ & $13.7465039696           $    & $ \quad  1.08\cdot 10^{-11} $
                                                              \\  \hline
$1$        & $1.00319691445           $    & $ -2.70\cdot 10^{-11} $
                                                              \\  \hline
$10$       & $6.155174431518\cdot 10^{-2} $& $ \quad  1.05\cdot 10^{-10} $
                                                              \\  \hline
$100$      & $3.83403357209\cdot 10^{-3} $ & $ -1.80\cdot 10^{-10} $
                                                              \\  \hline
$1000$     & $2.62958979905\cdot 10^{-4} $ & $ -1.04\cdot 10^{-12} $
                                                              \\  \hline
\end{tabular}
\end{table}

The second column in Table~\ref{tab:number3} was evaluated by program
QUADREC (see below) and also checked with the help of MATHEMATICA and MAPLE
V.

\section{Conclusion}
Thus it was shown that sufficiently complicated integrals (\ref{resid:3})
can be evaluated with a given accuracy by means of residue calculations and
further evaluation of series sum.

Evidently, that analytical-numeric method like this one, presented in this
paper, with some restrictions caused by the theory of the function of the
complex variable may be used for evaluation of a wide class of definite
integrals.

\section{Aknowledgements}
I am grateful Dr. V.K.Basenko and Dr. A.N.Berlisov for helpful discussions
and advice.
  
\appendix
\section{Appendix}
\subsection{Recursive adaptive quadrature program}
The algorithm consists of two practically independent parts: namely
adaptive procedure and quadrature formula.

The adaptive part uses effective recursive algorithm that implements
standard bisection method. To reach desired relative accuracy of the
integration the integral estimation over subinterval is compared with the
sum of integrals over two subintervals. If the accuracy is not reached the
adaptive part is called recursively (calls itself) for both (left and
right) subinterval.

Evaluation of integral sum on each step of bisection is performed by means
of quadrature formula. The construction of algorithm allows to choose which
type of quadrature to be used (should be used) throughout the integration.
Such possibility makes the code to be very flexible and applicable to a
wide range of integration problems.

Program realization of the described algorithm is possible when the
translator that (which) permits recursion is used. Here we propose a C++
version of such a recursive adaptive code. The text of the recursive
bisection function QUADREC (Quadrature used Adaptively and Recursively) is
presented below:
\vspace{-5mm}
\begin{tabbing}
\=\hspace{71mm}        \=\hspace{5cm} \\
\>{\small\tt void quadrec (TYPE x, TYPE X, TYPE whole) }\> \\
\>{\small\tt \{ static int recursion;            }\>{\small
                                   // Recursive calls counter} \\
\>{\small\tt if(++recursion < IP.recmax)         }\>{\small
                                   // Increase and check recursion level} \\
\>{\small\tt \{ TYPE section = (x+X)/2;          }\>{\small
                                   // Dividing the integration interval}  \\
\>{\small\tt TYPE left=IP.quadrature(x,section);}\>{\small
                                   // Integration over left subinterval}  \\
\>{\small\tt TYPE right=IP.quadrature(section,X);}\>{\small
                                   // Integration over right subinterval } \\
\>{\small\tt IP.result += left+right-whole;       } \>{\small
                                   // Modifying the integral value } \\
\>{\small\tt if((fabs(left+right-whole)>IP.epsilon*fabs(IP.result)))} \> \\
\>     \> // Checkup the accuracy \\
\>{\small\tt \{ quadrec(x,section,left);}     \> {\small
                                   // Recursion to the left subinterval}  \\
\>{\small\tt quadrec(section,X,right);\} \} }  \>{\small
                                   // Recursion to the right subinterval}  \\
\>{\small\tt else IP.rawint++;}              \>  {\small
                                   // Increase raw interval counter} \\
\>{\small\tt recursion--;}                   \>  {\small
                                   // Decrease recursion level counter}  \\
\>{\small\tt \}}                            \>
\end{tabbing}
\vspace{-2mm}

The form of the chosen comparison rule does not pretend on effectiveness
rather on simplicity and generality. Really it seems to be be very common
and does not depend on the integrand as well as quadrature type. At the
same time the use of this form in some cases result in overestimation of
the calculated integral consequently leads to more integrand function
calls. One certainly can get some gains, for instance, definite quadratures
with different number or/and equidistant points or Guass-Kronrod quadrature
etc.

Global structure IP used in the QUADREC function has to contain the
following fields:
\begin{verbatim}
struct iip {
TYPE (*fintegrand)(TYPE);     // Pointers on the integrand and
TYPE (*quadrature)(TYPE,TYPE);// quadrature function
TYPE epsilon;                 // Desired relative accuracy
int recmax;                   // Maximum number of recursions
TYPE result;                  // Result of the integration
int rawint;                   // Number of raw subintervals
} IP;
\end{verbatim}
The first four fields specify the input information and have to be defined
before calling the QUADREC function. The next two fields are the returned
values for the integration result and the number of unprocessed (raw)
subintervals. The static variable RECURSION in the QUADREC function is used
for controlling current recursion level. If the current level exceeds the
specified maximum number of recursions the RAWINT field is increased so
that its returned value will indicate the number of raw subintervals. Here
is the template for integration of function $f(x)$ over interval
$[X_{min},X_{max}]$ with the use of $q(x_1,x_2)$ quadrature formula:

\begin{verbatim}
IP.fintegrand = f;
IP.quadrature = q;
IP.epsilon = 1e-10;
IP.recmax = 30;
IP.result = IP.quadrature(Xmin,Xmax);
quadrec(Xmin,Xmax,IP.result);
\end{verbatim}

Crude estimation of the integral is evaluated by the quadrature function
and assigned to IP.result variable. Then it is transferred to QUADREC
function. Note that the initial estimation of the integral can be set in
principal to arbitrary value that usually does not alter the result of the
integration.

Further information about QUADREC: test results, using QUADREC in MS
Fortran source and so on, see \cite{BZh}.

\end{document}